\documentclass[12pt,a4paper]{article}
\usepackage{fullpage}
\usepackage[centertags]{amsmath}
\usepackage{amsbsy}
\usepackage{amssymb}
\allowdisplaybreaks[1]
\newcommand{\openone}{\leavevmode\hbox{\small1\kern-3.8pt\normalsize1}}
\begin{document}
\begin{flushright}
\begin{tabular}{l}
KIAS-P00070
\\
hep-ph/0011074
\\
November 6, 2000
\end{tabular}
\end{flushright}
\begin{center}
\Large\bfseries
Quantum Mechanics of Neutrino Oscillations
\\[0.5cm]
\normalsize\normalfont
Carlo Giunti
\\
\small\itshape
INFN, Sez. di Torino, and Dip. di Fisica Teorica,
Univ. di Torino, I--10125 Torino, Italy, and
\\
\small\itshape
School of Physics, Korea Institute for Advanced Study,
Seoul 130-012, Korea
\\[0.3cm]
\small\normalfont
and
\\[0.3cm]
\normalsize\normalfont
Chung W. Kim
\\
\small\itshape
School of Physics, Korea Institute for Advanced Study,
Seoul 130-012, Korea, and
\\
\small\itshape
Dept. of Physics $\&$ Astronomy,
The Johns Hopkins University,
Baltimore, MD 21218, USA
\end{center}
\begin{abstract}
We present a simple but general treatment of
neutrino oscillations in the framework of quantum mechanics
using plane waves and intuitive wave packet principles
when necessary.
We attempt to clarify some confusing statements
that have recently appeared in the literature.
\end{abstract}

\vspace{1cm}

The quantum mechanics of neutrino oscillations
have been studied in several papers
\cite{Bilenky-Pontecorvo-PR-78,%
Kayser-oscillations-81,%
Giunti-Kim-Lee-Whendo-91,%
Rich-93,%
Kiers-Nussinov-Weiss-PRD53-96,%
Campagne-97,%
Dolgov-Morozov-Okun-Shchepkin-97,%
Kiers-Weiss-PRD57-98,%
Giunti-Kim-Coherence-98,%
Nauenberg-Correlated-99,%
Takeuchi-Tazaki-Tsai-Yamazaki-00}
and reviewed in
\cite{CWKim-book,%
Zralek-oscillations-98}.
Alternative derivations of
neutrino oscillations
in the framework of quantum field theory
have also been presented in
\cite{Kobzarev-Martemyanov-Okun-Shchepkin-82,%
Giunti-Kim-Lee-Lee-93,%
Grimus-Stockinger-96,%
Shtanov-98,%
Giunti-Kim-Lee-Whendo-98,%
Ioannisian-Pilaftsis-solvable-98,%
Grimus-Mohanty-Stockinger-98,%
Cardall-Coherence-99,%
Grimus-Mohanty-Stockinger-99,%
Grimus-Mohanty-Stockinger-strength-99}.
In spite of all these studies, the understanding
of the theory of neutrino oscillations
seems still unsettled,
as one can see from controversial claims
in the recent literature
\cite{Grossman-Lipkin-spatial-97,%
Stodolsky-unnecessary-98,%
Srivastava-Charged-Lepton-95,%
Srivastava-Charged-Lepton-97,%
Srivastava-Widom-Sassaroli-Charged-98,%
Takeuchi-Tazaki-Tsai-Yamazaki-99,%
Rotelli-99}.

In this article we wish to clarify some of these issues
using a simple quantum mechanical treatment
of neutrinos as plane waves.
A rigorous treatment of neutrino oscillations
in the framework of quantum mechanics
requires a wave packet formalism
\cite{Kayser-oscillations-81}.
Here we employ some of its principles only when necessary,
without technical details
(see
\cite{Giunti-Kim-Lee-Whendo-91,%
Giunti-Kim-Coherence-98,%
CWKim-book,%
Zralek-oscillations-98}).

Neutrino oscillations are a consequence of neutrino mixing
and the fact that neutrino masses are very small.
Neutrino mixing implies that
the left-handed components
$\nu_{\alpha L}$
of the flavor neutrino fields
($\alpha=e,\mu,\tau$)\footnote{
We consider for simplicity
only the three active neutrino flavors
whose existence is firmly established
(see, for example, Ref.~\cite{PDG}).
However,
the formalism is valid for any number of neutrinos,
including additional sterile neutrinos
whose existence is under investigation
(see, for example, Refs.~\cite{BGG-review-98,Bilenky-Giunti-sterile-99}).

The flavor of an active neutrino
($\nu_e$, $\nu_\mu$, $\nu_\tau$)
is determined by the associated lepton
($e$, $\mu$, $\tau$)
in charged-current weak interactions.
Sterile neutrinos do not feel weak interactions,
as well as electromagnetic and strong interactions
(as active neutrinos);
they are sensitive only to gravitational interactions.
}
are superpositions of
the left-handed components
$\nu_{kL}$
of neutrino fields with definite mass $m_k$
($k=1,2,3$):
\begin{equation}
\nu_{\alpha L}
=
\sum_{k=1}^3
U_{\alpha k}
\,
\nu_{kL}
\qquad
(\alpha=e,\mu,\tau)
\,,
\label{0001}
\end{equation}
where $U$ is a unitary matrix
($U U^\dagger = U^\dagger U = \openone$),
\textit{i.e.} such that
\begin{align}
\null & \null
\sum_{k}
U_{\alpha k}
U_{\beta k}^*
=
\delta_{\alpha\beta}
\,,
\label{00021}
\\
\null & \null
\sum_{\alpha}
U_{\alpha k}^*
U_{\alpha j}
=
\delta_{kj}
\,.
\label{00022}
\end{align}

A neutrino with definite flavor is described
by the state
\begin{equation}
|\nu_\alpha\rangle
=
\sum_{k}
U_{\alpha k}^*
\,
|\nu_k\rangle
\,,
\label{0003}
\end{equation}
such that
\begin{equation}
\langle 0 | \, \nu_\alpha \, |\nu_\beta\rangle
=
\sum_{k,j}
U_{\alpha k}
U_{\beta j}^*
\,
\langle 0 | \, \nu_k \, |\nu_j\rangle
\propto
\sum_{k}
U_{\alpha k}
U_{\beta k}^*
=
\delta_{\alpha\beta}
\,,
\label{0004}
\end{equation}
where we used
$\langle 0 | \, \nu_k \, |\nu_j\rangle \propto \delta_{kj}$.
The state describing a flavor antineutrino
has complex-conjugated
elements of the mixing matrix with respect to the
flavor neutrino state
in Eq.~(\ref{0003}):
\begin{equation}
|\bar\nu_\alpha\rangle
=
\sum_{k}
U_{\alpha k}
\,
|\bar\nu_k\rangle
\,.
\label{00031}
\end{equation}
This implies that the oscillations
of neutrinos and antineutrinos are different only
if the mixing matrix is complex
(\textit{i.e.} only if there is CP violation).

Let us consider
$\nu_\alpha$
produced at time
$t_{\mathrm{P}}$
and space coordinate\footnote{
Since detected neutrinos propagate along
a macroscopic distance,
we consider only one spatial dimension along the neutrino path.
}
$x_{\mathrm{P}}$
by a weak interaction process
with a specific flavor $\alpha$
($\alpha=e,\mu,\tau$).
This neutrino is described by a state
$|\psi^{\mathrm{P}}_\alpha\rangle$
that is a superposition of
neutrino states with definite kinematical properties:
\begin{equation}
|\psi^{\mathrm{P}}_\alpha\rangle
=
\sum_{k}
U_{\alpha k}^*
\,
|\nu_k,p_k\rangle
\,.
\label{001}
\end{equation}
The state $|\nu_k,p_k\rangle$
describes a neutrino with mass $m_k$
and momentum $p_k$.
It is a product of the states
$|\nu_k\rangle$ and $|p_k\rangle$
belonging to the flavor-mass and momentum Hilbert spaces,
respectively:
\begin{equation}
|\nu_k,p_k\rangle
=
|\nu_k\rangle \, |p_k\rangle
\,.
\label{0011}
\end{equation}
In general the momenta of different mass eigenstates can
be different,
and they are determined by energy-momentum conservation
in the production process
\cite{Winter-81}.
For example,
in the case of
a muon neutrino produced by a pion decay at rest,
\begin{equation}
\pi^+ \to \mu^+ + \nu_{\mu}
\,,
\label{002}
\end{equation}
the momentum of the $k^{\mathrm{th}}$ mass eigenstate is
\begin{equation}
p_{k}^2
=
\frac{m_{\pi}^2}{4}
\left( 1 - \frac{m_{\mu}^2}{m_{\pi}^2} \right)^2
-
\frac{ m_{k}^2 }{ 2 }
\left( 1 + \frac{m_{\mu}^2}{m_{\pi}^2} \right)
+
\frac{ m_{k}^4 }{ 4 \, m_{\pi}^2 }
\,,
\label{0031}
\end{equation}
with the corresponding energy given by
\begin{equation}
E_{k}^2
=
\frac{m_{\pi}^2}{4}
\left( 1 - \frac{m_{\mu}^2}{m_{\pi}^2} \right)^2
+
\frac{ m_{k}^2 }{ 2 }
\left( 1 - \frac{m_{\mu}^2}{m_{\pi}^2} \right)
+
\frac{ m_{k}^4 }{ 4 \, m_{\pi}^2 }
\,,
\label{0032}
\end{equation}
where
$m_{\pi}$
and
$m_{\mu}$
are the masses
of the pion and muon, respectively.
The mass eigenstate
$|\nu_k,p_k\rangle$
satisfies the energy eigenvalue equation
\begin{equation}
\mathcal{H}
\,
|\nu_k,p_k\rangle
=
E_k
\,
|\nu_k,p_k\rangle
\,,
\label{005}
\end{equation}
where $\mathcal{H}$
is the free neutrino Hamiltonian.

In practice neutrinos with energy smaller than some fraction
of MeV are not detectable,
either because their energy is below threshold
in charged-current processes,
or because the cross section in elastic scattering processes,
which increases with neutrino energy,
is too small.
Hence,
if all the neutrino masses\footnote{
Direct searches of effective neutrino masses
in lepton decays
gave the upper limits
$m_{\nu_e} \lesssim 3 \, \mathrm{eV}$,
$m_{\nu_\mu} \lesssim 190 \, \mathrm{keV}$,
$m_{\nu_\tau} \lesssim 18.2 \, \mathrm{MeV}$
(see Ref.~\cite{PDG}).
However,
the sum of the masses of light neutrinos
is constrained to be smaller than about
$24 \, \mathrm{eV}$
by their contribution to the total energy density of the Universe
(see Ref.~\cite{PDG}).
}
are much smaller than 1 MeV,
all the detectable neutrinos
are extremely relativistic
and
it is convenient to
use a relativistic approximation
for the momentum and energy of the mass eigenstates.

At zeroth order in the relativistic approximation
all neutrino masses are considered negligible
and all the mass eigenstates have the same energy $E$,
equal to the modulus of their momentum:
$p_k = E_k = E$.
The value of $E$ is determined by energy-momentum conservation
in the production process.
For example,
from Eq.~(\ref{0031})
one can see that in the pion decay process in Eq.~(\ref{002})
we have
$
E
=
\frac{ m_{\pi} }{ 2 }
\left( 1 - \frac{ m_{\mu}^2 }{ m_{\pi}^2 } \right)
\simeq
30 \, \mathrm{MeV}
$.
Since the energy $E_k$ and momentum $p_k$
of the $k^{\mathrm{th}}$
mass eigenstate are related by the relativistic dispersion relation
\begin{equation}
E_k^2 = p_k^2 + m_k^2
\,,
\label{0060}
\end{equation}
the first order corrections to the equalities
$p_k = E_k = E$
depend on the square of the neutrino mass
and in general we can write
\begin{equation}
p_k
\simeq
E
-
\xi
\,
\frac{ m_{k}^2 }{ 2 E }
\,,
\label{0061}
\end{equation}
where $\xi$ is a quantity determined by energy-momentum
conservation in the production process.
For example,
from Eq.~(\ref{0031})
one can see that in pion decay
$
\xi
=
\frac{1}{2}
\left( 1 + \frac{m_\mu^2}{m_\pi^2} \right)
\simeq
0.8
$.
Using the relativistic approximation
of the energy-momentum dispersion relation in Eq.~(\ref{0060}),
\begin{equation}
E_k
\simeq
p_k
+
\frac{ m_{k}^2 }{ 2 E }
\,,
\label{00621}
\end{equation}
from Eq.~(\ref{0061})
we obtain that the energy $E_k$ of the mass eigenstate neutrino
$\nu_k$
at first order in the relativistic approximation
is given by
\begin{equation}
E_k
\simeq
E
+
( 1 - \xi )
\frac{ m_{k}^2 }{ 2 E }
\,.
\label{0062}
\end{equation}

The wave function corresponding to the initial flavor state
$|\psi^{\mathrm{P}}_\alpha\rangle$ in Eq.~(\ref{001})
is
\begin{equation}
|\psi^{\mathrm{P}}_\alpha(x)\rangle
=
\langle x|\psi^{\mathrm{P}}_\alpha\rangle
=
\sum_{k}
U_{\alpha k}^*
\,
\langle x|p_k\rangle |\nu_k\rangle
=
\sum_{k}
U_{\alpha k}^*
\,
e^{i p_k ( x - x_{\mathrm{P}} ) }
\,
|\nu_k\rangle
\,,
\label{007}
\end{equation}
where we have taken into account the initial coordinate
$x_{\mathrm{P}}$
of the production process.
The time evolution of the wave function
$|\psi^{\mathrm{P}}_\alpha(x)\rangle$
is given by the Schr\"odinger equation
\begin{equation}
i
\,
\frac{\mathrm{d}}{\mathrm{d}t}
\,
|\psi^{\mathrm{P}}_\alpha(x,t)\rangle
=
\mathcal{H}
\,
|\psi^{\mathrm{P}}_\alpha(x,t)\rangle
=
\sum_{k}
U_{\alpha k}^*
\,
e^{i p_k ( x - x_{\mathrm{P}} ) }
\,
E_k
\,
|\nu_k\rangle
\,,
\label{008}
\end{equation}
where we have used the energy eigenvalue equation (\ref{005}).
The solution of the evolution equation is
\begin{equation}
|\psi^{\mathrm{P}}_\alpha(x,t)\rangle
=
\sum_{k}
U_{\alpha k}^*
\,
e^{i p_k ( x - x_{\mathrm{P}} )
- i E_k ( t - t_{\mathrm{P}} ) }
\,
|\nu_k\rangle
\,,
\label{009}
\end{equation}
where $t_{\mathrm{P}}$ is the production time.
At a time different from the production time
$t_{\mathrm{P}}$
and a coordinate different from the production
coordinate
$x_{\mathrm{P}}$,
the state
$|\psi^{\mathrm{P}}_\alpha(x,t)\rangle$
is a superposition of flavor states.
Indeed,
expressing the mass eigenstate
$|\nu_k\rangle$
in Eq.~(\ref{009})
in terms of flavor eigenstates,
\begin{equation}
|\nu_k\rangle
=
\sum_{k}
U_{\alpha k}
\,
|\nu_\alpha\rangle
\label{010}
\end{equation}
(this relation is obtained by inverting Eq.~(\ref{0003})
using the unitarity relation in Eq.~(\ref{00022})),
one obtains
\begin{equation}
|\psi^{\mathrm{P}}_\alpha(x,t)\rangle
=
\sum_{\beta}
\left(
\sum_{k}
U_{\alpha k}^*
\,
e^{i p_k ( x - x_{\mathrm{P}} )
- i E_k ( t - t_{\mathrm{P}} ) }
\,
U_{\beta k}
\right)
|\nu_\beta\rangle
\,.
\label{011}
\end{equation}
This is a superposition of different flavor states,
which reduces to one flavor
only at the production point
$x=x_{\mathrm{P}}$, $t=t_{\mathrm{P}}$,
where the unitarity relation in Eq.~(\ref{00021})
guarantees that
$|\psi^{\mathrm{P}}_\alpha(x=x_{\mathrm{P}},t=t_{\mathrm{P}})\rangle = |\nu_\alpha\rangle$.

The term in parenthesis
in Eq.~(\ref{011})
is the amplitude of
$\nu_\alpha\to\nu_\beta$
transitions,
which is obtained by projecting
$|\psi^{\mathrm{P}}_\alpha(x=x_{\mathrm{D}},t=t_{\mathrm{D}})\rangle$
at the detection point space and time,
$x_{\mathrm{D}}$ and $t_{\mathrm{D}}$,
on the flavor state
$|\nu_\beta\rangle$:
\begin{equation}
\mathcal{A}_{\nu_\alpha\to\nu_\beta}(L,T)
=
\langle\nu_\beta|\psi^{\mathrm{P}}_\alpha(x=x_{\mathrm{D}},t=t_{\mathrm{D}})\rangle
=
\sum_{k}
U_{\alpha k}^*
\,
e^{i p_k L - i E_k T }
\,
U_{\beta k}
\,,
\label{012}
\end{equation}
where
$L \equiv x_{\mathrm{D}}-x_{\mathrm{P}}$
is the source-detector distance
and
$T \equiv t_{\mathrm{D}}-t_{\mathrm{P}}$
is the neutrino propagation time.

Here it is to be emphasized that the detection process
is described only by the flavor state
$|\nu_\beta\rangle$,
with no need to express the detected neutrino state
as a superposition of mass eigenstates with different momenta.
This is due to the principle of causality,
which implies that
different mass eigenstates with different momenta
created in the production process
determine their own kinematics in the detection process
where they are detected.
The corresponding different momenta of the particles
participating to the detection process
do not spoil the coherence of the process
if they are within the momentum uncertainty
due to the localization of the detection process
(through the uncertainty principle).

The probability
of 
$\nu_\alpha\to\nu_\beta$
transitions is
\begin{equation}
P_{\nu_\alpha\to\nu_\beta}(L,T)
=
|\mathcal{A}_{\nu_\alpha\to\nu_\beta}(L,T)|^2
\,,
\label{013}
\end{equation}
leading to
\begin{equation}
P_{\nu_\alpha\to\nu_\beta}(L,T)
=
\sum_k
|U_{\alpha k}|^2
|U_{\beta k}|^2
+
2
\mathrm{Re}
\sum_{k>j}
U_{\alpha k}^*
U_{\beta k}
U_{\alpha j}
U_{\beta j}^*
\,
e^{i (p_k-p_j) L - i (E_k-E_j) T }
\,.
\label{0141}
\end{equation}
It is clear that the transition probability at the detector
depends only on the neutrino mixing matrix and on the
momentum and energy differences of the mass eigenstates,
which are due to their mass differences.
Invoking now the relativistic approximations
in Eqs.~(\ref{0061}) and (\ref{0062}),
we obtain
\begin{align}
P_{\nu_\alpha\to\nu_\beta}(L,T)
=
\null & \null
\sum_k
|U_{\alpha k}|^2
|U_{\beta k}|^2
\nonumber
\\
\null & \null
+
2
\mathrm{Re}
\sum_{k>j}
U_{\alpha k}^*
U_{\beta k}
U_{\alpha j}
U_{\beta j}^*
\,
\exp\left[
- i \xi \, \frac{\Delta{m}^2_{kj}}{2E} \, L
- i \left(1-\xi\right) \, \frac{\Delta{m}^2_{kj}}{2E} \, T
\right]
\,,
\label{014}
\end{align}
with the mass-squared differences
$\Delta{m}^2_{kj} \equiv m_k^2 - m_j^2$.

Equation (\ref{014})
gives the oscillation probability as a function
of the distance $L$ between a source and a detector
and the neutrino travel time $T$.
However,
in real experiments,
the neutrino travel time is not measured,
but the source-detector distance is known.
Therefore,
the neutrino travel time $T$
must be expressed in terms of the distance $L$.
To accomplish this task it is necessary to
go beyond the plane wave approximation employed so far
and remind that the propagating mass eigenstate neutrinos
must be described by wave packets.
Indeed, a plane wave has an infinite space-time extent,
whereas neutrino production and detection are localized processes,
that are appropriately described by the wave packet formalism.
We associate each mass eigenstate,
which has well defined kinematical properties,
with a wave packet.
The spatial width of the neutrino wave-packets
is determined by the maximum between the
quantum uncertainties of the position and time
of the microscopic production process
(see Ref.~\cite{Giunti-Kim-Coherence-98}).
The wave packets propagate with the group velocity
\begin{equation}
v_k
=
\frac{p_k}{E_k}
\simeq
1 - \frac{m_k^2}{2E^2}
\,,
\label{015}
\end{equation}
where we have used the relativistic approximations
(\ref{0061}) and (\ref{0062}).
The different mass eigenstates can interfere
if they overlap upon their arrival at the detector
or if the coherence time of the detection process
is larger than the time separation between the mass eigenstates
\cite{Kiers-Nussinov-Weiss-PRD53-96,%
Kiers-Weiss-PRD57-98,%
Giunti-Kim-Coherence-98}.
In any case,
from Eq.~(\ref{015}) one can see that
the arrival time at the detector
of the mass eigenstate wave packets
is equal to the source-detector distance\footnote{
We use natural units in which $c=\hbar=1$.
}
plus a correction proportional to
$m_k^2/E^2$.
Since the phase in the neutrino oscillation probability
in Eq.~(\ref{014})
is already proportional to
$\Delta{m}^2_{kj}/E$,
the correction to the relation $T=L$
gives a contribution of higher order,
which can be neglected.
One may object that since this correction is also proportional to
the distance $L$,
it is not negligible when
$(\Delta{m}^2_{kj}L/E)(m_k^2/E^2) \gtrsim 1$.
This is true.
However,
one must observe that in practice neutrino oscillations
are observable only when the
phase $(\Delta{m}^2_{kj}L/E)$ is of order one.
This is due to the fact that all
neutrino beams have a spread in energy,
all detectors have a finite energy resolution
and the source-detector distance
has a macroscopic uncertainty due to the
sizes of the source and detector.
When the oscillation phase
is much larger than one,
the averages over the neutrino energy
and the source-detector distance
wash out the oscillating term.
If all the phases are large and all the oscillating terms are washed out,
there remains only a constant
transition probability given by the first term in Eq.~(\ref{014}):
\begin{equation}
\langle P_{\nu_\alpha\to\nu_\beta} \rangle
=
\sum_k
|U_{\alpha k}|^2
|U_{\beta k}|^2
\,.
\label{016}
\end{equation}
Therefore,
the observability of neutrino oscillations
guarantees that
$(\Delta{m}^2_{kj}L/E)(m_k^2/E^2) \ll 1$
and the correction to the relation $T=L$
is negligible.

Another correction to the relation $T=L$
is due to the finite time interval
in which the wave packets
overlap with the detection process
and to the finite coherence time of the detection process.
One should average the oscillation probability in Eq.~(\ref{014})
over the resulting total time interval $\Delta{t}_{\mathrm{D}}$.
However,
since the time interval $\Delta{t}_{\mathrm{D}}$
is characteristic of a microscopic process,
it is always much smaller than the macroscopic propagation time
$T \simeq L$.
If this property were not satisfied,
oscillations could not be observed because
they would be washed out by the average over time.
As mentioned above,
since neutrino oscillations are observable
only when the
phase $(\Delta{m}^2_{kj}T/E)$ is of order one,
the correction
$(\Delta{m}^2_{kj}\Delta{t}_{\mathrm{D}}/E) \ll 1$
is always negligible.
In other words,
the phase of the oscillations
is practically constant in the time interval $\Delta{t}_{\mathrm{D}}$
and the average over such a time interval
is equivalent to the approximation $T=L$.

From Eq.~(\ref{014}),
the appropriate approximation $T=L$ leads to the
transition probability
\begin{equation}
P_{\nu_\alpha\to\nu_\beta}(L)
=
\sum_k
|U_{\alpha k}|^2
|U_{\beta k}|^2
+
2
\mathrm{Re}
\sum_{k>j}
U_{\alpha k}^*
U_{\beta k}
U_{\alpha j}
U_{\beta j}^*
\,
\exp\left[
- i \frac{\Delta{m}^2_{kj}}{2E} \, L
\right]
\,,
\label{017}
\end{equation}
as a function of the source-detector distance $L$ measured in real experiments.
This is the standard oscillation formula derived
more than 20 years ago
\cite{Eliezer-Swift-MIXING-76,%
Fritzsch-Minkowski-OSCILLATIONS-76,%
Bilenky-Pontecorvo-AGAIN-76,%
Bilenky-Pontecorvo-PR-78}.
Notice that as a result of the $T=L$ relation,
$\xi$ has magically disappeared,
leading to the independence of the neutrino oscillation probability
from the specific details of the production process.
From Eq.~(\ref{017})
one can see that there is an oscillating term
associated with each pair of mass eigenstates
$k>j$
and the corresponding oscillation length is
\begin{equation}
L_{kj}
=
\frac{4\pi E}{\Delta{m}^2_{kj}}
\,.
\label{0171}
\end{equation}

Since the state in Eq.~(\ref{00031}) describing a flavor antineutrino
has complex-conjugated
elements of the mixing matrix with respect to the
flavor neutrino state,
the probability of
$\bar\nu_\alpha\to\bar\nu_\beta$
oscillations is given by Eq.~(\ref{017})
with complex-conjugated
elements of the mixing matrix.
The oscillation lengths of neutrinos and antineutrinos are equal.

Some comments are in order for the clarification
of the derivation of the neutrino
oscillation probability
in Eq.~(\ref{017}):

\renewcommand{\labelenumi}{\arabic{enumi})}
\begin{enumerate}

\item
\label{item10}
The initial neutrino state in Eq.~(\ref{001})
is a \emph{coherent}
superposition of mass eigenstates.
This is possible because neutrino masses are small
and
the differences between the momenta
of different mass eigenstates
is smaller than the momentum uncertainty
of the production process
due to its localization in space.
Denoting the spatial interval of the source by
$\Delta{x}_{\mathrm{P}}$,
we have the momentum uncertainty
$\Delta{p}_{\mathrm{P}} \sim 1/\Delta{x}_{\mathrm{P}}$
and the two mass eigenstates
$\nu_k$, $\nu_j$
are emitted coherently
if
$|p_k - p_j| \lesssim \Delta{p}_{\mathrm{P}}$.
Using Eq.~(\ref{0061})
one gets the condition
$|\Delta{m}^2_{kj}| \lesssim 2E\Delta{p}_{\mathrm{P}}/\xi$.
For example,
assuming
$\Delta{x}_{\mathrm{P}} \lesssim 1 \, \mathrm{cm}$
in the process in Eq.~(\ref{002}) of pion decay at rest,
we obtain the coherence condition
$|\Delta{m}^2_{kj}| \lesssim 10^2 \, \mathrm{eV}^2$,
which is well satisfied by neutrino masses below eV ranges.

If the momentum uncertainty in the production process
is too small,
different mass eigenstates are emitted incoherently
by the source and neutrinos do not oscillate.
This happens if the momenta of the particles in the production process
are measured with such a high accuracy to the extent that one can determine
which neutrino mass eigenstate is produced.
In this case
$
\Delta{x}_{\mathrm{P}}
\gtrsim
2E/\xi|\Delta{m}^2_{kj}|
\sim
L^{\mathrm{osc}}_{kj}
$,
\textit{i.e.}
the uncertainty in the localization of the source is larger
than the oscillation length
\cite{Kayser-oscillations-81}.
It is obvious that in this case it is not possible to measure
neutrino oscillations.
Analogous considerations
apply to the detection process.

\item
\label{item11}
We have assumed that all the mass eigenstates
are produced at the same space-time point
$(x_{\mathrm{P}},t_{\mathrm{P}})$
and detected at the same
space-time point
$(x_{\mathrm{D}},t_{\mathrm{D}})$.
Since both the production and detection processes
have uncertainties in space and time,
different mass eigenstates can be produced and detected coherently
at different locations in space and time
within the uncertainties,
generating interferences responsible for neutrino oscillations.
However,
one must pay special attention to the
\emph{coherent} character of the production and detection processes.
In other words,
there is a well-defined phase relation
between the wave functions of
neutrinos produced (detected) within the spatial and temporal coherence intervals
of the production (detection) process.

Let us first consider each mass eigenstate neutrino $\nu_k$
produced at a different space-time point
$(x_{\mathrm{P}}^{k},t_{\mathrm{P}}^{k})$,
with
$|x_{\mathrm{P}}^{k}-x_{\mathrm{P}}| \lesssim \Delta{x}_{\mathrm{P}}$
and
$|t_{\mathrm{P}}^{k}-t_{\mathrm{P}}| \lesssim \Delta{t}_{\mathrm{P}}$.
Since the component of the neutrino field that creates
the corresponding mass eigenstate in the production process
oscillates in space-time with a phase determined by the
energy $E_k$ and momentum $p_k$,
the initial phase difference between
the wave function of the $k^{\mathrm{th}}$
mass eigenstate neutrino produced at
$(x_{\mathrm{P}}^{k},t_{\mathrm{P}}^{k})$
and that
produced at
$(x_{\mathrm{P}},t_{\mathrm{P}})$
is
\begin{equation}
\Delta\phi_{\mathrm{P}}^{k}
=
p_k \left( x_{\mathrm{P}}^{k}-x_{\mathrm{P}} \right)
-
E_k \left( t_{\mathrm{P}}^{k}-t_{\mathrm{P}} \right)
\,.
\label{018}
\end{equation}
Then,
the phase at $(x,t)$ of
the wave function of the $k^{\mathrm{th}}$
mass eigenstate neutrino produced at
$x_{\mathrm{P}}^{k}$
and time
$t_{\mathrm{P}}^{k}$
is
\begin{equation}
\phi_k(x,t)
=
\Delta\phi_{\mathrm{P}}^{k}
+ p_k ( x - x_{\mathrm{P}}^{k} )
- E_k ( t - t_{\mathrm{P}}^{k} )
=
p_k ( x - x_{\mathrm{P}} )
- E_k ( t - t_{\mathrm{P}} )
\,,
\label{019}
\end{equation}
\textit{i.e.}
the same as that
of
the wave function of the $k^{\mathrm{th}}$
mass eigenstate neutrino produced at
$x_{\mathrm{P}}$
and time
$t_{\mathrm{P}}$,
used in Eq.~(\ref{009}).

Let us consider now
each mass eigenstate neutrino $\nu_k$
detected at a corresponding space-time point
$(x_{\mathrm{D}}^{k},t_{\mathrm{D}}^{k})$,
with
$|x_{\mathrm{D}}^{k}-x_{\mathrm{D}}| \lesssim \Delta{x}_{\mathrm{D}}$
and
$|t_{\mathrm{D}}^{k}-t_{\mathrm{D}}| \lesssim \Delta{t}_{\mathrm{D}}$.
The incoming neutrino wavefunction is
\begin{equation}
|\psi^{\mathrm{P}}_\alpha(x_{\mathrm{D}}^{k},t_{\mathrm{D}}^{k})\rangle
=
\sum_{k}
U_{\alpha k}^*
\,
e^{i p_k ( x_{\mathrm{D}}^{k} - x_{\mathrm{P}} )
- i E_k ( t_{\mathrm{D}}^{k} - t_{\mathrm{P}} ) }
\,
|\nu_k\rangle
\,.
\label{0231}
\end{equation}
Since the component of the neutrino field $\nu_k$ that destroys
the corresponding mass eigenstate in the detection process
oscillates in space-time with a phase determined by the
energy $E_k$ and momentum $p_k$,
its phase difference between
the points
$(x_{\mathrm{D}}^{k},t_{\mathrm{D}}^{k})$
and
$(x_{\mathrm{D}},t_{\mathrm{D}})$
is
\begin{equation}
\Delta\phi_{\mathrm{D}}^{k}
=
p_k \left( x_{\mathrm{D}}^{k}-x_{\mathrm{D}} \right)
-
E_k \left( t_{\mathrm{D}}^{k}-t_{\mathrm{D}} \right)
\,.
\label{0181}
\end{equation}
This implies that
the detected neutrino in a $\nu_\alpha\to\nu_\beta$
oscillation experiment
is no longer described by a simple flavor state $|\nu_\beta\rangle$
as in Eq.~(\ref{012}),
but by the state
\begin{equation}
|\psi^{\mathrm{D}}_\beta(x_{\mathrm{D}}^{k},t_{\mathrm{D}}^{k})\rangle
=
\sum_k
U_{\beta k}^*
\,
e^{i \Delta\phi_{\mathrm{D}}^{k}}
\,
|\nu_k\rangle
\,,
\label{023}
\end{equation}
which takes into account the coherence of the process.
The amplitude of $\nu_\alpha\to\nu_\beta$
transitions
turns out to be
\begin{equation}
\mathcal{A}_{\nu_\alpha\to\nu_\beta}
=
\langle\psi^{\mathrm{D}}_\beta(x_{\mathrm{D}}^{k},t_{\mathrm{D}}^{k})|
\psi^{\mathrm{P}}_\alpha(x_{\mathrm{D}}^{k},t_{\mathrm{D}}^{k})\rangle
=
\sum_{k}
U_{\alpha k}^*
\,
e^{i p_k L - i E_k T }
\,
U_{\beta k}
\,,
\label{024}
\end{equation}
which is equal to the standard amplitude in Eq.~(\ref{012}).
This mechanism allows the interference
of different mass eigenstates even when the corresponding
wave packets do no overlap,
but arrive at the detector within the
coherence time interval of the detection process
\cite{Kiers-Nussinov-Weiss-PRD53-96,%
Kiers-Weiss-PRD57-98,%
Giunti-Kim-Coherence-98}.
Clearly,
in this case the different mass eigenstates are
detected at different times,
but the coherence of the detection process
allows them to interfere.

\item
\label{item12}
In the simple plane wave approach that
we have adopted above,
there is no indication of a coherence length
for neutrino oscillations \cite{Nussinov-coherence-76}.
The reason for the existence of a coherence length
is that
each massive neutrino propagates as a wave packet with its own group velocity.
The overlap of different wave packets decreases
with increasing distance from the source,
until eventually they separate.
For distances larger than the coherence length,
two wave packets arrive at the detector
separated by a time interval larger than the coherence time
of the detection process.
In this case
the corresponding mass eigenstates cannot interfere
and the oscillations are suppressed.
Hence,
in general
there is a coherence length associated with each pair of mass eigenstates.
At distances larger than the
largest coherence length
one can only measure the averaged constant probability
in Eq.~(\ref{016}).
This can be calculated using the wave packet formalism
in the framework of quantum mechanics
\cite{Giunti-Kim-Lee-Whendo-91,Giunti-Kim-Coherence-98}
or quantum field theory
\cite{Giunti-Kim-Lee-Lee-93,Giunti-Kim-Lee-Whendo-98}.
However,
one can estimate the coherence length in the following simple way,
as done in Ref.~\cite{Nussinov-coherence-76}.
Using the group velocity formula in Eq.~(\ref{015}),
the separation of the
$k^{\mathrm{th}}$
and
$j^{\mathrm{th}}$
mass eigenstate wave packets
after traveling a time $T$ corresponding to a distance $L$ is
\begin{equation}
|\Delta{x}_{kj}|
=
\left| v_k - v_j \right| T
\simeq
\frac{|\Delta{m}^2_{kj}|}{2E} L
\,.
\label{021}
\end{equation}
The size of the mass eigenstate wave packets
is given by the maximum between 
the spatial and temporal coherence widths of the
production process.
Denoting by $\sigma_x$
the maximum between the size of the mass eigenstate wave packets
and
the spatial and temporal coherence widths of the
detection process,
the interference is suppressed for
$|\Delta{x}_{kj}| \gtrsim \sigma_x$,
leading to the coherence length
\begin{equation}
L_{kj}^{\mathrm{coh}}
\sim
\frac{2E\sigma_x}{|\Delta{m}^2_{kj}|}
\,.
\label{022}
\end{equation}
For $n$ massive neutrinos
there are $n(n-1)/2$ oscillating terms in the probability
in Eq.~(\ref{017}),
each one with its oscillation length in Eq.~(\ref{0171})
and coherence length in Eq.~(\ref{022}).

\end{enumerate}

Several controversial and often misleading statements and derivations
of the neutrino oscillation formula have recently appeared
in the literature.
Let us comment on some of them:

\renewcommand{\labelenumi}{\Alph{enumi})}
\begin{enumerate}

\item
\label{item21}
The traditional derivation of the neutrino oscillation probability
\cite{Bilenky-Pontecorvo-PR-78}
is based on the assumption that all the mass eigenstates have a common momentum
$p_k=p_j=E$.
This is often called ``equal momentum''
treatment of neutrino oscillations.
From Eq.~(\ref{0061})
one can see that the equal momentum assumption
is equivalent to assuming $\xi=0$,
\textit{i.e.}
this case is simply
a particular case of the general
kinematical relations in Eqs.~(\ref{0061}) and (\ref{0062}).
Since the derivation of the neutrino oscillation formula in Eq.~(\ref{017})
does not depend on the value of $\xi$,
it is clear that the equal momentum assumption
is an acceptable one.

\item
\label{item22}
It has been argued
\cite{Grossman-Lipkin-spatial-97}
that neutrinos created by a definite weak-interaction process
must have a definite flavor at the production point
at all times.
From Eq.~(\ref{009})
one can see that this is possible only if all the mass eigenstates
have the same energy,
$E_k=E_j=E$.
Based on this considerations,
the authors of Ref.~\cite{Grossman-Lipkin-spatial-97}
claim that the ``equal energy'' assumption is the correct one.
However,
since the production process has uncertainties both in time and space,
one could claim as well that
neutrinos created by a definite weak-interaction process
must have a definite flavor over all
the spatial region of the production point
at an initial time
(or during the coherence time).
From Eq.~(\ref{009})
one can see that this is possible only if all the mass eigenstates
have the same momentum,
$p_k=p_j=E$,
justifying
the equal momentum assumption.
Since energy and momentum satisfy the dispersion relation
$E_k^2 = p_k^2 + m_k^2$,
different mass eigenstates
cannot have simultaneously the same energy and momentum.
Hence, it is clear that
the equal energy assumption is as arbitrary as the
equal momentum assumption,
being yet another special case,
$\xi=1$,
of the general
kinematical relations in Eqs.~(\ref{0061}) and (\ref{0062}).
Obviously,
the equal energy assumption
leads to the same oscillation formula
in Eq.~(\ref{017})
that we have derived without arbitrary assumptions.

\item
\label{item23}
It has been claimed that there is a
``factor of two ambiguity''
in the oscillation lengths
\cite{Grossman-Lipkin-spatial-97,Rotelli-99}.
This ambiguity arises from the observation
that
the propagation time $T$
in Eq.~(\ref{0141})
is not common among the
different mass eigenstates,
but one should take a different propagation time
$T_k$
for each mass eigenstate,
given by
$T_k = L / v_k$,
where $v_k$
is the group velocity
in Eq.~(\ref{015}).
In this case the phase in the $k$--$j$ interference term is
$
\phi_{kj}
=
(p_k-p_j) L - (E_k T_k-E_j T_j)
$.
In the relativistic approximation
$
E_k T_k
\simeq
p_k L
+
m_k^2 L / E
$,
leading to
$
\phi_{kj}
=
\Delta{m}^2_{kj} L / E
$.
One can see that the phases are twice as large as the standard ones in Eq.~(\ref{017}),
giving oscillation lengths that are a half of the standard ones in Eq.~(\ref{0171}).
This is the claimed factor of two ambiguity\footnote{
A similar ambiguity has been claimed to exist in the case of kaon oscillations
in Refs.~\cite{Lipkin-nonexperiments-95,%
Srivastava-Widom-Sassaroli-Lambda-95,%
Srivastava-Widom-Sassaroli-correlations-95,%
Srivastava-Widom-Lambda-96}.
The authors of Refs.~\cite{Lowe-Lambda-96,Kayser-QM-97}
tried to solve this problem arguing that
the interference between the $K_L$ and $K_S$ components
should be calculated at the same time and
the same space point.
However,
this argument does not hold because,
as discussed in Ref.~\cite{Kiers-Weiss-PRD57-98},
the coherent character of the detection process
for a finite space-time interval allows
wave functions at different space-times
to interfere
(see item \ref{item11} above).
}.
The confusion in this derivation is that
it does not take into account the fact that
different mass eigenstates arriving at different times at the detector
must be absorbed \emph{coherently} by the detection process
in order to interfere.
Hence,
there is a phase relation of the detection process
that must be taken into account,
as discussed in item \ref{item11} above.
This restores the missing factor of two in the oscillation lengths.

\item
\label{item24}
As a variant of the equal momentum and equal energy assumptions,
a ``equal velocity'' assumption has been proposed
\cite{Takeuchi-Tazaki-Tsai-Yamazaki-99,%
Takeuchi-Tazaki-Tsai-Yamazaki-00,%
Rotelli-99}.
Simple kinematics does not allow this possibility
\cite{Okun-Tsukerman-00}.
Indeed,
$v_k = v_j$
implies that
$p_k/E_k = p_j/E_j$.
Taking the square of this relation
and using the energy-momentum dispersion relation
$p_k^2 = E_k^2 - m_k^2$,
one finds
$m_k/m_j = E_k/E_j$.
This equality cannot be true in a real experiment,
since
$E_k/E_j \simeq 1$,
whereas
$m_k/m_j$
may be extremely small or large.

\item
It has been argued that a wave packet treatment of neutrino oscillations
is unnecessary when neutrinos are emitted by a source that
is stationary in time \cite{Stodolsky-unnecessary-98}.
In this argument there is confusion between microscopic and macroscopic
stationarity.
In practice most neutrino sources
(as the sun, or a reactor, or a supernova)
can be considered to be stationary sources of neutrinos.
However,
the microscopic processes that generate neutrinos
in these sources cannot be considered stationary,
because
each microscopic process is able to emit neutrino waves
only for a finite interval of time,
the coherence time that we have denoted above by
$\Delta{t}_{\mathrm{P}}$.
Hence,
to understand the physics of neutrino oscillations,
a wave packet treatment \emph{is} necessary.
We have seen above that
a wave packet description of the propagating
mass eigenstates is crucial
for the determination of the correct relation
between the neutrino propagation time and the source-detector distance.
The necessity to establish such a connection
is avoided in Ref.~\cite{Stodolsky-unnecessary-98}
by assuming that all mass eigenstates have the same energy.
However,
we have shown in item \ref{item22} above
that in general such assumption is not justified
(the stationarity argument given in
Ref.~\cite{Stodolsky-unnecessary-98}
obviously does not apply to the microscopic
production process).

\item
In neutrino oscillation experiments
neutrinos are usually produced through charged-current
weak processes together with a charged lepton
(see, for example, Eq.~(\ref{002})).
It is natural to ask if also the charged lepton
has some kind of oscillatory behavior.
In principle,
flavor oscillations of charged leptons
would be possible
if their masses were very close so as
to be produced coherently,
and if there were some other means,
besides measurement of mass,
to distinguish different charged lepton flavors.
However,
since
these two requirements are not satisfied in the real world,
there are no flavor oscillations of charged leptons.
Notice, in particular,
that charged leptons are defined by their mass,
which is the only property that distinguishes among
charged lepton flavors\footnote{
In practice all detectors reveal particles through electromagnetic interactions.
Charged leptons have all the same electromagnetic interactions.
They are distinguished either through kinematics
or through their decay products.
Both these characteristics depend directly from the charged lepton mass.
}.
In other words,
flavor is measured through mass.
That is why charged leptons
are defined as mass eigenstates.

In Refs.~\cite{Srivastava-Charged-Lepton-95,%
Srivastava-Charged-Lepton-97,%
Srivastava-Widom-Sassaroli-Charged-98}
it has been claimed that the probability to detect the charged lepton
oscillates in space-time\footnote{
A similar effect,
called ``Lambda oscillations'',
has been claimed to exist
\cite{Srivastava-Widom-Sassaroli-Lambda-95,%
Srivastava-Widom-Lambda-96}
for the $\Lambda$'s
produced together with a neutral kaon,
as in the process
$ \pi^- + p \to \Lambda + K^0 $.
This effect has been refuted in
Refs.~\cite{Lowe-Lambda-96,%
Lowe-recoil-98}.
}.
This claim has been refuted in
Ref.~\cite{Dolgov-Morozov-Okun-Shchepkin-97}.
The authors of Ref.~\cite{Srivastava-Charged-Lepton-95}
presented a proof that the oscillation in space-time
of the probability to detect the charged lepton
is a consequence of the oscillation in space-time
of the probability to detect the neutrino.
But it is well known that
the probability to detect the neutrino,
irrespective of its flavor,
does not oscillate in space-time.
This property is usually called ``conservation of probability''
or ``unitarity''
and is represented mathematically by the general relation
$
\sum_\beta
P_{\nu_\alpha\to\nu_\beta}
=
1
$.
Hence,
the argument presented in Ref.~\cite{Srivastava-Charged-Lepton-95}
actually proves that charged leptons
do \emph{not} oscillate!

\end{enumerate}

In conclusion,
we have presented a simple but general
treatment of neutrino oscillations
in a quantum mechanical framework
using plane waves
and intuitive wave packets arguments when necessary.
We have shown that the
standard neutrino oscillation formula in Eq.~(\ref{017})
can be derived without arbitrary assumptions
as ``equal momentum'' or ``equal energy''.
We also tried to
clarify some confusing
statements on the quantum mechanics of neutrino oscillations
that have recently appeared in the literature.

\begin{flushleft}
\large\textbf{Acknowledgments}
\end{flushleft}
C.G. would like to thank P. Rotelli
for stimulating discussions during the
NOW2000 workshop.
C.G. would also like to express his gratitude to
the Korea Institute for Advanced Study (KIAS)
for warm hospitality.


\end{document}